\documentclass[twocolumn]{jpsj2}
\usepackage{revsymb}
\usepackage{bm}
\newcommand{\ket}[1]{|{#1}\rangle}
\newcommand{\bra}[1]{\langle{#1}|}
\newcommand{\Tr}{\mathop{\text{Tr}}\nolimits}
\renewcommand{\Re}{\mathop{\text{Re}}\nolimits}

\def\runauthor{K. \textsc{Yuasa} \textit{et~al.}}
\title{%
Purification of Quantum State through Zeno-like Measurements}
\author{%
Kazuya \textsc{Yuasa},\thanks{E-mail address: yuasa@hep.phys.waseda.ac.jp} Hiromichi \textsc{Nakazato}\thanks{E-mail address: hiromici@waseda.jp} and Tomoko \textsc{Takazawa}}
\inst{%
Department of Physics, Waseda University, Tokyo 169-8555, Japan}
\recdate{February 14, 2003}
\abst{%
We present a novel procedure to purify quantum states, i.e., purification through Zeno-like measurements.
By simply repeating one and the same measurement on a quantum system, one can purify another system in interaction with the former.
The conditions for the (efficient) purification are specified on a rather general setting, and the framework of the method possesses wide applicability.
It is explicitly demonstrated on a specific setup that the purification becomes very efficient by tuning relevant parameters.
}
\kword{purification of quantum states, quantum Zeno effect, repeated measurement}
\begin{document}
\maketitle
One cannot observe a quantum system without any disturbance on the object, and the dynamics of the quantum system is affected by the measurement.
An interesting manifestation of this effect is the ``quantum Zeno effect'' (QZE)~\cite{ref:ClassicQZE,ref:QZEMisraSudarshan,ref:QZEReview}: 
if one repeats measurement very frequently in order to ascertain whether a system remains in the initial state, the evolution of the system is slowed down and totally hindered in the limit of infinite frequency.
Or conversely, one can also accelerate the decay of an unstable state by repeating the measurement less frequently than a critical frequency (inverse QZE)~\cite{ref:IQZE}. 
These are typical examples of the effects of measurement on quantum dynamics.

In this article, we present another interesting effect of measurement on quantum dynamics when the measurement is repeated as in the case of the (inverse) QZE: \textit{purification of quantum state through Zeno-like measurements}.~\cite{ref:qpf}
We show in the following that a series of measurements on a quantum system indirectly affects the dynamics of another system in interaction with the former, and the latter (initially in any mixed state) is driven into a \textit{pure} state, provided certain conditions are satisfied.
The state of the latter system is purified through the repeated measurement on the former.

In the ideas of quantum information and computation, quantum coherence plays crucial roles~\cite{ref:QInfoCompZeilinger,ref:QInfoCompNielsen}.
It is hence one of the important issues to be addressed how to maintain and/or recover quantum coherence, and various schemes have been proposed attacking this subject~\cite{ref:QInfoCompZeilinger,ref:QInfoCompNielsen,ref:Purification,ref:PurificationCirac,ref:ErrorCorrection,ref:DecoherenceFreeSubspaces,ref:ZenoSubspace,ref:DynamicalDecoupling,ref:TasakiPascazio}. 
The work presented here contributes to this issue and provides a novel method to purify quantum states (i.e., to recover quantum coherence), which is simple compared to the standard purification techniques~\cite{ref:QInfoCompZeilinger,ref:Purification,ref:PurificationCirac} and the framework is rather general.

Let a total system $\text{A}+\text{B}$ be described by a Hamiltonian of the form
\begin{equation}
H=H_\text{A}+H_\text{B}+H_\text{int},
\label{eqn:Hamiltonian}
\end{equation}
where $H_\text{A(B)}$ is a free Hamiltonian of system A(B) and $H_\text{int}$ is an interaction between A and B\@.
We prepare system A in a pure state $\ket{\phi}$ at time $t=0$, while the initial state of system B, denoted by $\rho_\text{B}$, can be arbitrary (a mixed state).
The total system starts to evolve from the initial state
\begin{equation}
\rho_0=\ket{\phi}\bra{\phi}\otimes\rho_\text{B}
\end{equation}
with the Hamiltonian~(\ref{eqn:Hamiltonian}), and we check the state of system A at time $t=\tau$.
If it is confirmed that system A remains in its initial state $\ket{\phi}$, the state of the total system is projected by a projection operator
\begin{equation}
\mathcal{O}=\ket{\phi}\bra{\phi}\otimes\openone
\label{eqn:ProjectionOp}
\end{equation}
and restarts to evolve.
We repeat such a measurement regularly at time intervals $\tau$ (Zeno-like measurement) as long as system A is found in the state $\ket{\phi}$ at every step.

Notice that we do not check the state of system B and the projection operator~(\ref{eqn:ProjectionOp}) does not return the time-evolved total system to its initial state: system A is set back to its initial state $\ket{\phi}$ but system B is not.
System B evolves away from its initial state under the sequence of measurements on system A\@.
We are interested in such evolution of system B, which is indirectly affected by the measurement repeated on system A through the interaction $H_\text{int}$.
We shall show in the following that such evolution can result in a purification phenomenon: system B evolves into a \textit{pure} state irrespectively of its initial (mixed) state $\rho_\text{B}$, if certain conditions specified below are satisfied.

System A is confirmed to be in the state $\ket{\phi}$ successively $N$ times with the probability
\begin{align}
P^{(\tau)}(N)
&=\Tr[(\mathcal{O}e^{-iH\tau}\mathcal{O})^N\rho_0
(\mathcal{O}e^{iH\tau}\mathcal{O})^N]\nonumber\\
&=\Tr_\text{B}[\bm{(}V_\phi(\tau)\bm{)}^N\rho_\text{B}
\bm{(}V_\phi^\dag(\tau)\bm{)}^N],
\label{eqn:Probability}
\end{align}
where the operator $V_\phi(\tau)\equiv\bra{\phi}e^{-iH\tau}\ket{\phi}$ is an operator acting on the Hilbert space of system B\@.
(Once system A is found in a different state from $\ket{\phi}$, we stop proceeding to the next step, failing to purify the state of system B\@.)
After the $N$ successful confirmations, the total and B systems are in the states
\begin{align}
\rho^{(\tau)}(N)
&=\ket{\phi}\bra{\phi}\otimes\rho_\text{B}^{(\tau)}(N),\\
\rho_\text{B}^{(\tau)}(N)
&=\bm{(}V_\phi(\tau)\bm{)}^N\rho_\text{B}\bm{(}V_\phi^\dag(\tau)\bm{)}^N/P^{(\tau)}(N),
\label{eqn:StateB}
\end{align}
respectively.

In the ordinary situation where the measurement is repeated at an infinite frequency by taking the limit $N\to\infty$ keeping $T=N\tau$ finite, the probability $P^{(\tau)}(N)$ in~(\ref{eqn:Probability}) increases as $N$ becomes large, approaching unity $P^{(\tau)}(N)\to1$ and the ordinary QZE~\cite{ref:QZEMisraSudarshan} appears.
At the same time, the dynamics of system B in~(\ref{eqn:StateB}) becomes unitary $\bm{(}V_\phi(T/N)\bm{)}^N\to\mathcal{V}_\phi(T)$ (a unitary operator) in this limit.
This is an example of the so-called ``quantum Zeno dynamics.''~\cite{ref:QZD,ref:ZenoSubspace,ref:TasakiPascazio}
We are however interested in a different situation: we repeat the measurement with a nonvanishing (not necessarily small) $\tau$.
The probability $P^{(\tau)}(N)$ would decay out completely for such a finite $\tau$, but we are interested in the asymptotic behavior of the state of system B, $\rho_\text{B}^{(\tau)}(N)$, for large (but finite) values of $N$\@.

In order to clarify the evolution of system B under the Zeno-like measurement on system A, let us consider the eigenvalue problem of the projected time-evolution operator $V_\phi(\tau)$.
Since the operator $V_\phi(\tau)$ is not Hermitian, we need to set up both the right- and left-eigenvalue problems
\begin{equation}
V_\phi(\tau)|u_n)=\lambda_n|u_n),\quad
(v_n|V_\phi(\tau)=\lambda_n(v_n|.
\end{equation}
The eigenvalue $\lambda_n$ is in general complex valued and satisfies $0\le|\lambda_n|\le1$, $\forall n$, which reflects the unitarity of the time-evolution operator.
Let us assume for the moment that the spectrum of the operator $V_\phi(\tau)$ is discrete and nondegenerate.
The eigenvectors then form an orthonormal complete set in the following sense:
\begin{equation}
\sum_n|u_n)(v_n|=\openone,\quad
(v_m|u_n)=\delta_{mn}.
\end{equation}
The operator $V_\phi(\tau)$ itself is expanded in terms of the eigenvectors
\begin{equation}
V_\phi(\tau)=\sum_n\lambda_n|u_n)(v_n|,
\end{equation}
and we obtain
\begin{equation}
\bm{(}V_\phi(\tau)\bm{)}^N=\sum_n\lambda_n^N|u_n)(v_n|,
\end{equation}
which asymptotically behaves for large $N$ as
\begin{equation}
\bm{(}V_\phi(\tau)\bm{)}^N\xrightarrow{\text{large}\ N}
\lambda_0^N|u_0)(v_0|
\label{eqn:Mechanism}
\end{equation}
\textit{if the largest (in magnitude) eigenvalue $\lambda_0$ is unique, discrete, and nondegenerate}.
It is now evident that the state of system B, $\rho_\text{B}^{(\tau)}(N)$ given in~(\ref{eqn:StateB}), asymptotically approaches a pure state
\begin{equation}
\rho_\text{B}^{(\tau)}(N)
\xrightarrow{\text{large}\ N}|u_0)(u_0|/(u_0|u_0),
\end{equation}
i.e., as we repeat the confirmation that system A is in the state $\ket{\phi}$ at regular intervals $\tau$, system B is driven into the pure state $|u_0)$.
This is what we call ``purification through Zeno-like measurements.''

Notice that the final pure state $|u_0)$ is independent of the initial state of system B, $\rho_\text{B}$, i.e., any initial (mixed) state is purified into the unique pure state $|u_0)$ through the Zeno-like measurement on system A\@.
The pure state $|u_0)$ is the eigenstate belonging to the largest (in magnitude) eigenvalue $\lambda_0$ of the projected time-evolution operator $V_\phi(\tau)$ and is prescribed by the Hamiltonian $H$, the measurement interval $\tau$, and the state $\ket{\phi}$ onto which system A is repeatedly projected.
We therefore have the possibility to purify system B into a \textit{desired} pure state by adjusting $\tau$, $\ket{\phi}$, and $H$.

But one cannot always purify system B successfully.
System A should be found in the state $\ket{\phi}$ at \textit{every} measurement for the purification to be accomplished, but the probability for this to occur, which is nothing but the probability $P^{(\tau)}(N)$ given in~(\ref{eqn:Probability}), is less than $1$, unfortunately.
It decays asymptotically as
\begin{equation}
P^{(\tau)}(N)
\xrightarrow{\text{large}\ N}|\lambda_0|^{2N}(u_0|u_0)(v_0|\rho_\text{B}|v_0),
\label{eqn:Pasymp}
\end{equation}
which gives the yield of the purification protocol presented here.
For an efficient purification with high yield $P^{(\tau)}(N)$, you can adjust parameters again to satisfy $|\lambda_0|=1$, which prevents the yield $P^{(\tau)}(N)$ from decaying out completely.
We will come back to this point below.

Before going on, let us recapitulate the conditions for the purification.
The heart of the mechanism of the purification is the asymptotic behavior of the operator $\bm{(}V_\phi(\tau)\bm{)}^N$ in~(\ref{eqn:Mechanism}), which is true \textit{if the largest (in magnitude) eigenvalue $\lambda_0$ is unique, discrete, and nondegenerate}.
Although we assumed above for concreteness that the spectrum of the operator $V_\phi(\tau)$ is discrete and nondegenerate, these assumptions on the spectrum is not essential for the purification except for the conditions on the largest (in magnitude) eigenvalue $\lambda_0$.
One has only to check the largest (in magnitude) eigenvalue $\lambda_0$, which should be \textit{unique, discrete, and nondegenerate}.
In the field of quantum information and computation, for example, finite level (two- or three-level) systems play important roles, and we can easily expect that such systems with discrete spectra fulfill these conditions.

Let us illustrate the purification described above with a model Hamiltonian
\begin{equation}
H=\Omega a^\dag a+\omega b^\dag b+ig(a^\dag b-ab^\dag),
\label{eqn:Oscillator}
\end{equation}
i.e., two single-mode harmonic oscillators $a$ and $b$ in interaction with the rotating-wave approximation.
The frequencies $\Omega$ and $\omega$ and the coupling constant $g$ are real parameters.
At time $t=0$, we prepare oscillator $a$ in a coherent state $\ket{\alpha}$ specified by a complex parameter $\alpha$, while the state of oscillator $b$ is arbitrary.
We let the total system evolve with the Hamiltonian~(\ref{eqn:Oscillator}) and confirm repeatedly at regular intervals $\tau$ that oscillator $a$ is in the coherent state $\ket{\alpha}$.
The time-evolution operator (between adjacent measurements), $e^{-iH\tau}$, is calculated exactly to be
\begin{equation}
e^{-iH\tau}=e^{Aa^\dag b}e^{Ba^\dag a}e^{Cb^\dag b}e^{-Aab^\dag}
\end{equation}
in terms of the $\tau$-dependent functions
\begin{subequations}
\begin{align}
A&=\frac{(g/\delta)\sin\delta\tau}{\cos\delta\tau+i[(\Omega-\omega)/2\delta]\sin\delta\tau},\\
B&=-\frac{i}{2}(\Omega+\omega)\tau-\ln\!\left(
\cos\delta\tau+i\frac{\Omega-\omega}{2\delta}\sin\delta\tau
\right),\\
C&=-\frac{i}{2}(\Omega+\omega)\tau+\ln\!\left(
\cos\delta\tau+i\frac{\Omega-\omega}{2\delta}\sin\delta\tau
\right),
\end{align}
\end{subequations}
where $\delta=\sqrt{g^2+(\Omega-\omega)^2/4}$, and the projected time-evolution operator $V_\alpha(\tau)\equiv\bra{\alpha}e^{-iH\tau}\ket{\alpha}$ reads
\begin{subequations}
\begin{equation}
V_\alpha(\tau)
=e^{-|\alpha|^2[1-e^B-A^2/(1-e^{-C})]}e^{D(b^\dag,b)},
\end{equation}
\begin{equation}
D(b^\dag,b)=C\left(
b^\dag+\frac{A\alpha^*}{1-e^{-C}}
\right)\left(
b-\frac{A\alpha}{1-e^{-C}}
\right),
\end{equation}
\end{subequations}
whose spectrum is given by
\begin{subequations}
\begin{equation}
\lambda_n=e^{-|\alpha|^2[1-e^B-A^2/(1-e^{-C})]}e^{nC},
\end{equation}
\begin{equation}
|u_n)=U|n),\quad
(v_n|=(n|U^{-1},
\end{equation}
\begin{equation}
U=\exp\!\left[
\frac{A}{1-e^{-C}}(\alpha^*b+\alpha b^\dag)
\right],
\end{equation}
\end{subequations}
where $n=0,1,2,\ldots$ and $\{|n)\}$ are the number states of oscillator $b$.
Since the absolute value of $e^C$ is $|e^C|=\sqrt{1-(g/\delta)^2\sin^2\!\delta\tau}\le1$, the largest (in magnitude) eigenvalue of the relevant operator $V_\alpha(\tau)$ is $\lambda_0$ unless $|e^C|=1$, which is explicitly given by
\begin{align}
\lambda_0&=e^{-|\alpha|^2[1-e^B-A^2/(1-e^{-C})]}\nonumber\\
&=\exp\bigglb[
-2|\alpha|^2\biggl\{
1-\frac{i}{2}\,\biggl[
\left(1+\frac{\Omega-\omega}{2\delta}\right)
\cot\frac{\Omega_+\tau}{2}\nonumber\\
&{}\phantom{{}=\exp\bigglb[-2|\alpha|^2\biggl\{1-{}\,}
+\left(1-\frac{\Omega-\omega}{2\delta}\right)
\cot\frac{\Omega_-\tau}{2}
\biggr]
\biggr\}^{-1}
\biggrb],
\label{eqn:Lambda0Model}
\end{align}
where $\Omega_\pm=(\Omega+\omega)/2\pm\delta$, and is discrete and nondegenerate, satisfying the conditions for the purification.
We see therefore that oscillator $b$, initially in any (mixed) state, is purified into the pure state $|u_0)$, i.e., into a coherent state $|\tilde{\alpha})$
\begin{equation}
|u_0)=U|0)=|\tilde{\alpha})\equiv\left|\frac{A\alpha}{1-e^{-C}}\right),
\label{eqn:FinalCoherentState}
\end{equation}
as we repeatedly confirm at time intervals $\tau$ that oscillator $a$ is in the coherent state $\ket{\alpha}$, provided $\delta\tau\neq m\pi$ $(m=0,1,2,\ldots)$.

In fact, for an initial state $\rho_\text{B}\propto e^{-\beta\omega b^\dag b}$ ($\beta$ is a positive parameter), the state of oscillator $b$ under the Zeno-like measurement, $\rho_\text{B}^{(\tau)}(N)$, is explicitly evaluated to be
\begin{subequations}
\begin{equation}
\begin{split}
\rho_\text{B}^{(\tau)}(N)
{}={}&(1-|e^C|^{2N}e^{-\beta\omega})\\
&{}\times D(N)e^{-(\beta\omega-2N\Re C)b^\dag b}D^\dag(N),
\end{split}
\end{equation}
\begin{equation}
D(N)=\exp\!\left(
-\frac{\alpha\Theta(N)b^\dag-\alpha^*\Theta^*(N)b}{1-|e^C|^{2N}e^{-\beta\omega}}
\right),
\end{equation}
\begin{equation}
\Theta(N)
=\frac{1-e^{NC}}{1-e^{-C}}A
-\left(\frac{1-e^{NC}}{1-e^{-C}}A\right)^*e^{NC}e^{-\beta\omega},
\end{equation}
\end{subequations}
which indeed approaches the coherent state $|\tilde{\alpha})$ in~(\ref{eqn:FinalCoherentState}),
\begin{equation}
\rho_\text{B}^{(\tau)}(N)
\xrightarrow{N\to\infty}D(\infty)|0)(0|D^\dag(\infty)
=|\tilde{\alpha})(\tilde{\alpha}|,
\end{equation}
if $\Re C\neq0$.
[$|0)$ here is normalized to unity, i.e., $(0|0)=1$, and also $(\tilde{\alpha}|\tilde{\alpha})=1$.]

Let us finally discuss the efficiency of the purification.
For this purpose, we introduce the so-called fidelity $F^{(\tau)}(N)$ for the target pure state $|u_0)$:
\begin{equation}
F^{(\tau)}(N)\equiv(u_0|\rho_\text{B}^{(\tau)}(N)|u_0).
\end{equation}
We want to achieve high fidelity $F^{(\tau)}(N)$ close to $1$ with high yield $P^{(\tau)}(N)$ after a small number of measurements, $N$.
We can make the fidelity $F^{(\tau)}(N)$ as high as we want by indefinitely increasing the number of measurements, $N$, while the yield $P^{(\tau)}(N)$ would keep decaying asymptotically as~(\ref{eqn:Pasymp}).
It is however possible to prevent the decay of the yield $P^{(\tau)}(N)$ in~(\ref{eqn:Pasymp}) by adjusting relevant parameters to satisfy the condition
\begin{equation}
|\lambda_0|=1.
\label{eqn:ConditionI}
\end{equation}
If, at the same time, the absolute values of the other eigenvalues $\lambda_n$ ($\forall n\neq0$) are much smaller than that of $\lambda_0$, i.e.,
\begin{equation}
|\lambda_n/\lambda_0|\ll1\quad(\forall n\neq0),
\label{eqn:ConditionII}
\end{equation}
we can purify a quantum state quickly.
Equations~(\ref{eqn:ConditionI}) and~(\ref{eqn:ConditionII}) are the conditions for an efficient purification.

\begin{figure}[b]
\begin{center}
\includegraphics[width=0.45\textwidth]{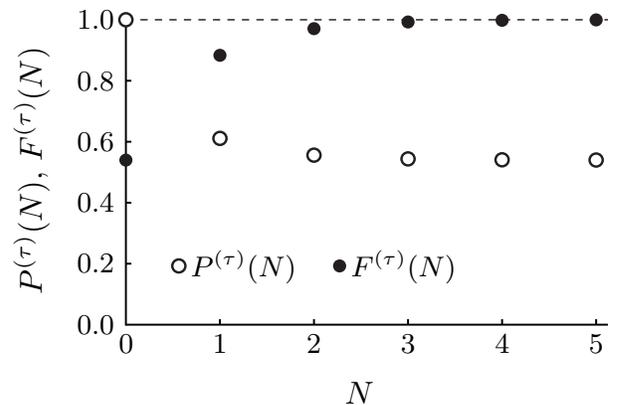}
\end{center}
\caption{Fidelity $F^{(\tau)}(N)$ and yield $P^{(\tau)}(N)$ for the oscillator model~(\ref{eqn:Oscillator}) when the initial state of oscillator $a$, onto which $N$ measurements are performed, is a coherent state $\ket{\alpha}$ and that of oscillator $b$ is $\rho_\text{B}\propto e^{-\beta\omega}$.
The parameters are taken to be $\omega=1$, $g=0.2$, $\beta=1$, $\alpha=0.5$, and $\tau=2\pi/\Omega_+\simeq5.24$ in units such that $\hbar=\Omega=1$.
$\tau$ is tuned so as to satisfy the condition~(\ref{eqn:ConditionI}) and the ratio of the second largest (in magnitude) eigenvalue $\lambda_1$ to the largest one $\lambda_0$ is $|\lambda_1/\lambda_0|=|e^C|\simeq0.37$.}
\label{fig:FidelityYield}
\end{figure}

For the above oscillator model, we can indeed achieve these conditions: by tuning the time interval between measurements as $\tau=2m\pi/|\Omega_\pm|$ ($m=0,1,2,\ldots$), we have $\lambda_0=1$.
[See~(\ref{eqn:Lambda0Model}).]
The fidelity $F^{(\tau)}(N)$ and the yield $P^{(\tau)}(N)$ for the oscillator model are shown in Fig.~\ref{fig:FidelityYield}, where $\tau$ is tuned to be $\tau=2\pi/\Omega_+$.
The decay of the yield $P^{(\tau)}(N)$ is suppressed due to this tuning, and a very high fidelity is achieved after only $N=2$ steps since the ratio of the second largest (in magnitude) eigenvalue $\lambda_1$ to the largest one $\lambda_0$, i.e., $|\lambda_1/\lambda_0|=|e^C|$,
is well less than $1$.
(See the caption of Fig.~\ref{fig:FidelityYield}.)

The above discussions clearly reveal that repeated measurements on a quantum system dramatically affect the dynamics of another system in interaction with the former and the latter is driven into a pure state irrespectively of its initial (mixed) state.
One can purify quantum states through Zeno-like measurements.
This procedure is very simple compared to the other standard techniques~\cite{ref:QInfoCompZeilinger,ref:Purification,ref:PurificationCirac}: we have only to repeat the same measurement.
One can prescribe the final pure state by tuning the measurement interval $\tau$, the state $\ket{\phi}$ onto which the former (controlled) system is repeatedly projected, and the Hamiltonian $H$.

Since we have clarified the mechanism of the purification under a rather general setup, without specifying the Hamiltonian, it possesses wide applicability.
The only conditions for the purification one should check is that the largest (in magnitude) eigenvalue $\lambda_0$ of the projected time-evolution operator $V_\phi(\tau)$ be unique, discrete, and nondegenerate.
Furthermore, careful control of this eigenvalue $\lambda_0$ leads us to an efficient purification, the conditions for which are given in~(\ref{eqn:ConditionI}) and~(\ref{eqn:ConditionII}).

In the field of quantum information and computation, entangled states play important roles, and many people are involved with the issue of ``entanglement purification/distillation.''~\cite{ref:QInfoCompZeilinger,ref:Purification}
Our general framework of the purification also applies to this issue, and one can purify quantum entanglement through Zeno-like measurements discussed here, details of which will be reported elsewhere~\cite{ref:Plenio}.
Further studies, e.g., to clarify the optimality of this method~\cite{ref:PurificationCirac}, to propose some realistic physical setups, and so on, would provide us with deeper understandings on the present method.

The authors acknowledge fruitful discussions with Professor I.\ Ohba.
K.Y.\ thanks Professors S. Tasaki and I. Ojima for comments and encouragements, and Professor T. Petrosky for discussions at the symposium.
This work is partly supported by Grants-in-Aid for Scientific Research (C) from the Japan Society for the Promotion of Science (No.~14540280) and Priority Areas Research (B) from the Ministry of Education, Culture, Sports, Science and Technology, Japan (No.~13135221), and by a Waseda University Grant for Special Research Projects (No.~2002A-567).

\end{document}